\documentclass[usegraphicx,useAMS,usenatbib]{mn2e}
\title[DDRGS as remnants of binary black holes]{Double-double radio galaxies: 
       remnants of merger of supermassive binary black holes}
\author[F.K. Liu et al]{F.K. Liu$^{1,2}$\thanks{E-mail: fkliu@bac.pku.edu.cn 
        (FKL), wuxb@bac.pku.edu.cn (XBW)}, Xue-Bing Wu$^{1}$ and
		S.L. Cao$^{3}$ \\
$^1$ National Astronomical Observatory of Chinese Academy of Sciences \&
	Astronomy Department of Peking University,\\
	100871 Beijing, China; \\
$^2$Department of Astronomy, G\"oteborg University \& Chalmers University 
	of Technology, 41296 G\"oteborg, Sweden \\
$^3$Astronomy Department, Beijing Normal University, 100875 Beijing China
}

\begin{document}

\date{Accepted ***.
      Received ***;
      in original form ***}

\pagerange{\pageref{firstpage}--\pageref{lastpage}} \pubyear{1994}

\maketitle

\label{firstpage}

\begin{abstract}

      The activity of active galaxy may be triggered by 
the merge of galaxies and present-day galaxies are probably the product 
of successive minor mergers. The frequent galactic merges at high redshift
imply that active galaxy harbors supermassive unequal-mass binary black holes 
in its center at least once during its life time. The secondary black hole 
interacts and becomes coplanar with the accretion disk around the primary,
inspiraling toward their mass center due to the loss of the orbit angular
momentum to the disk mass outside the orbit of the secondary and/or to the 
gravitational radiation. The binary black holes finally merge and form a 
more massive (post-merged) black hole at center. In this paper, we 
showed that the recently discovered double-lobed FR II radio galaxies 
are the remnants of such supermassive binary black holes. The 
inspiraling secondary black hole opens a gap in the accretion disk, which
increases with time when the loss of the orbit angular momentum via 
gravitational radiation becomes dominated. When the supermassive black 
holes merge, inner accretion disk disappears and the gap becomes a big hole
of about several hundreds of Schwarzschild radius in the vicinity of the 
post-merged supermassive black hole, leading to an interruption of jet 
formation. When the outer accretion disk slowly refills the big hole on a 
viscous time scale, the jet formation restarts and the interaction of the 
recurrent jets and the inter-galactic medium forms a secondary pair of 
lobes. We applied the model to a particular double-lobed radio source 
B1834+620, which has an interruption time scale $\sim 1\, {\rm Myr}$. We
showed that the orbit of the secondary in B1834+620 is elliptical with a
typical eccentricity $e \simeq 0.68$ and the mass ratio $q$ of the secondary
and the primary is $ 0.01 \la q \la 0.4$. The accretion disk is a standard
$\alpha$-disk with $0.01 \la \alpha \la 0.04$ and the ratio  
of disk half height $H$ and radius $r$ is $\delta \simeq 0.01$. The model 
predicates that double-lobed radio structure forms only in FR II or 
borderline FR I/FR II radio galaxies and the detection rate of double-lobed
radio sources among FRII radio sources is about one percent. 

\end{abstract}

%-----

%-----
\section{Introduction}

      Active galactic nuclei (AGNs) consists of a super-massive black 
hole surrounded by an accretion disk, continuously supplying energy of the
extended radio lobes via narrow and relativistic plasma jets. Prominent 
and continuous large scale extra-galactic radio jets have been clearly 
detected in 661 radio sources \citep{liu02}. Among the extra-galactic 
radio sources, about ten FR II  radio galaxies \citep{fanaroff74} are 
very peculiar and consist of a pair of symmetric double-lobed radio 
structures with one common center and two extended and edge-brightened 
inner radio lobes \citep{schoenmakers00a,schoenmakers00b,saripalli02}.  
The inner structure has a well aligned axis with the outer lobes and a
relatively lower luminosity. These radio sources are called double-double
radio galaxies (DDRGs) and their structures are most likely due to the 
interruption and restarting of jet formation at the central engine with 
an interruption time of the order of Myr \citep{schoenmakers00a}. 
Interruption- and recurrent-jet phenomenon is also detected in some 
non-DDRG radio sources, e.g. 3C288 \citep{bridle89}, 3C219
\citep{clarke92}, B1144+352 \citep{schoenmakers99}, and the
compact symmetric object (CSO) B0108+388 \citep{baum90,owsianik98}.

      While the evolution of re-current jets in the intergalactic medium 
(IGM) has been investigated in some detail \citep{clarke91,reynolds97,
kaiser00}, the mechanism to interrupt and restart the jet formation in the
center of AGNs is unclear. The proposed scenarios in literature include a
passive magnetic field model \citep{clarke92}, internal instability in the
accretion disk due to the radiation pressure induced warping \citep{pringle97,
natarajan98}, a large fraction of the gas left by the secondary galaxy in 
a merging or colliding galaxy system \citep{schoenmakers00a}. However, the 
passive magnetic field model is not consistent with the observations of the
DDRG source B1834+620 \citep{lara99}. The the internal warping instability
is likely to change the direction of the jet considerably \citep{natarajan98}.
The falling or colliding gas model has difficulty to explain the abrupt
interruption and restarting of jet formation and is not consistent with the 
observations of no evidence for galaxy interaction in DDRGs. 

      It was found recently that the central black hole masses in active and 
inactive galaxies have tight correlations with the central velocity dispersions
(e.g. \citet{gebhardt00,merritt01a,merritt01b}) and the bulge luminosities 
(e.g. \citet{magorrian98,mclure02}) of the host galaxies. These relations 
imply that the activity of AGNs may be triggered by galaxy merging and 
present-day galaxies are probably the product of successive minor mergers
\citep{kauffmann00,haehnelt00,menou01}. The frequent galactic mergers at 
high red-shift imply that active galaxy harbors super-massive unequal-mass
binary black holes at center at least once in its life time. Supermassive 
binary black holes may have been observed, e.g. in the BL Lac object OJ287
\citep{sillanpaa88,liu02b}. Once the super-massive binary black holes form,
the secondary interacts with the gas in the circumbinary accretion disk and 
becomes coplanar, sinking towards the mass center and getting merged due to 
the loss of the orbit angular momentum to the disk mass outside the 
orbit and/or to the gravitational radiation \citep{goldreich80,lin86,
pringle91,artymowicz92,artymowicz94,syer95,scheuer96,ivanov99,narayan00,
gould00,armitage02,zhao02}. We show in this paper that DDRGs are the 
remnants of the binary-disk interactions and the coalescence of supermassive
binary black holes. As the interaction between the secondary and an
advection-dominated accretion flow (ADAF) is negligible \citep{narayan00}.
we consider only a standard thin $\alpha$-disk \citep{shakura73} or a slim
disk \citep{abramowicz88} and assume that the rotating primary black
hole aligns with the accretion disk due to the Lense-Thirring effect
\citep{scheuer96,natarajan98}. We consider an 
elliptical binary system of an initial semimajor axis $a \sim 10^3 
r_{\rm G}$, where the shrink of the binary separation is driven by the
viscous loss of the angular momentum of the secondary to the outer 
accretion disk. Here $r_{\rm G}$ is the Schwarzschild radius of the 
primary black hole. As it takes very long time for the secondary to 
pass through the region of $a \sim 10^5 r_{\rm G}$ to $10^3 r_{\rm G}$
\citep{begelman80,quinlan97,ivanov99}, the binary system is an old system.
As the secondary-disk interaction always tends to align the disk and the
orbital plane of the secondary \citep{scheuer96,vokrouhlicky98,ivanov99},
we assume that the orbit plane of the binary and the disk are 
coplanar at $a \sim 10^3 r_{\rm G}$. The secondary interacts radially
with the accretion disk and opens a gap in it. When the gravitation 
radiation dominates the loss of the orbital angular momentum at a 
smaller semimajor axis $a$, the secondary rapidly pushes inwards the 
gas trapped inside the orbit and the gap gets wider with the decreasing
of the binary separation. When two super-massive black holes merge,
the gap becomes a big hole in the vicinity of the primary and and the inner
accretion disk disappears. The jet formation interrupts. When the inner 
edge of the outer accretion disk slowly involves inwards and reaches the 
last stable orbit, the big hole is refilled with the disk materials and jet 
formation restarts. We show that the observed interruption time of jet
formation in DDRGs is the viscous time for the accreted plasma to refill
the inner disk. This model can also give explanations to many other 
observations of DDRGs.

      We describe our model in Sec.~\ref{sec:model}. The application to
DDRGs, in particular B1834+620, is given in Sec.~\ref{sec:appl}. Our
discussions and conclusions are presented in Sec.~\ref{sec:con}.

%------------------------------------

\section[]{Creation of a big hole in the accretion disk}
\label{sec:model}

      If the orbit of the secondary is coplanar with the accretion disk 
and the mass ratio $q = m/M$ of the secondary and the primary is 
\begin{equation}
1 \gg q > q_{\rm min} = {81 \pi \over 8} \alpha \delta^2 \simeq 3 
        \times 10^{-5} \alpha_{-2} \delta_{-2}^2 , 
\label{eq:llimit}
\end{equation}
the secondary black hole opens a gap in the disk and exchanges angular 
momentum with disk gas via gravitational torques \citep{lin86}. 
In Eq.~\ref{eq:llimit}, $\alpha = 0.01 \alpha_{-2}$ is the viscous
parameter, $\delta = 0.01 \delta_{-2} = H/r $ and $H$ is the half 
thickness of the disk. For a gas-pressure dominated accretion disk, 
$\delta$ is nearly independent of radius $r$ \citep{collin90}
\begin{eqnarray}
\delta & = & {H \over r} \nonumber \\
       &\simeq& 0.01 \alpha_{-2}^{-1/10} 
         \left({L \over 0.1 L_{\rm E}}\right)^{1/5} M_8^{-1/10}
	\left({\epsilon \over 0.2}\right)^{-1/5} \times
	\nonumber \\
	& &\left({r \over 10^3 r_{\rm G}}\right)^{1/20} ,
\label{eq:delta}
\end{eqnarray}
where $L_{\rm E} = 6.9 \times 10^{46} M_8 \ {\rm erg\, s^{-1}}$ with 
$M_8 = M /(5 \times 10^8 M_\odot)$ is the Eddington luminosity, 
$\epsilon = L / \dot{M} c^2$ is the efficiency of the accretion process
and can be as high as $0.4$ for a Kerr black hole. Eq.~\ref{eq:delta}
implies that $\delta$ is insensitive to all the variables and we will
take 0.01 as its standard value. If the total disk mass $M_{\rm d}$ 
inside the disk radius $r_{\rm d}$ is $M_{\rm d} \ga m$ and the 
separation $a$ of the binary is large, the secondary migrates inwards 
on the viscous time at a speed \citep{lin86,syer95,ivanov99}
\begin{equation}
  \dot{a}_{\rm vis} \simeq -{3 \over 2} {\nu \over r} 
	\simeq - {3 \over 2} \delta^2 \alpha v_{\rm K} ,
\label{eq:avis}
\end{equation}
where $v_{\rm K}$ is the Keplerian velocity and $ H/r \simeq c_{\rm s} 
/ v_{\rm K}$. When $a$ is small, the loss of the angular 
momentum due to gravitational radiation becomes important. At some 
critical radius $a_{\rm cr}$, the migration speed $\dot{a}_{\rm vis}$
is comparable to the inspiraling rate due to the gravitational radiation 
\citep{peters63}
\begin{equation}
    \dot{a}_{\rm gw} = - {64 G^3 M^3 q \left(1 + q\right) \over 5
        c^5 a^3 } f = - {8\over 5} \left({r_{\rm G} \over a}\right)^3 q 
	\left(1 + q\right) f c,
\label{eq:agw}
\end{equation}
where $f$ is a function of the eccentricity $e$
\begin{equation}
f = \left(1 + {73 \over 24} e^2 + {37 \over 96} e^4\right) 
	\left(1 - e^2\right)^{-7/2} .
\label{eq:fec}
\end{equation}
The orbit of the secondary is circular due to the binary-disk interaction
for a binary system of $q \la 10^{-2}$ but is elliptical for $q \ga 
10^{-2}$ \citep{artymowicz92}. As minor galactic mergers are 
more common than major mergers in the hierarchical models of galaxy 
formation and $10^{-2} \la q \ll 1 $ \citep{haehnelt00},
the eccentricity is in the range of $0 < e \la 0.75$. Therefore, 
Eq.~\ref{eq:llimit} is always satisfied for reasonable values of $\alpha$
and $\delta$ and the secondary opens a gap in the accretion disk. From
Eqs.~\ref{eq:avis} and \ref{eq:agw}, we have
\begin{equation}
    a_{\rm cr} = {1 \over 2} \left({128 \over 15}\right)^{2/5} 
               \delta^{-4/5} \alpha^{-2/5} q^{2/5} \left(1 + 
	       q\right)^{1/5} f^{2/5} r_G .
\label{eq:acr}
\end{equation}
When $a = a_{\rm cr}$, the inner edge of the accretion disk outside the
orbit of the secondary (outer disk) is at $r_{\rm o} \simeq n^{2/3} 
a_{\rm cr}$ and the outer edge of the accretion disk inside the orbit
(inner disk) is at $r_{\rm i} \simeq n^{-2/3} a_{\rm cr}$ 
\citep{artymowicz94}, where $n$ is the resonance number: $n = 2$ for
a circular orbit and $n = 5$ for $\alpha \sim 0.01$ and $e \sim 0.5$. 
For convenience, we define a critical time scale $t_{\rm cr} \equiv 
a_{\rm cr} / |\dot{a}_{\rm gw}| = a_{\rm cr} / |\dot{a}_{\rm vis}|$ 
and
\begin{equation}
t_{\rm cr} = {1 \over 3} \left({128 \over 15}\right)^{3/5} 
        \delta^{-16/5} \alpha^{-8/5} \left({q^3 \over 1 +
	  q}\right)^{1/5}
	f^{3/5} \left({r_{\rm G} \over c}\right) .
\label{eq:tcr}
\end{equation}
For a typical disk-binary system of
$\alpha = 0.01$, $\delta = 0.01$ and $M= 5 \times 10^8 {\rm M_\odot}$, we 
have $a_{\rm cr} \simeq 110 r_{\rm G}$ and $t_{\rm cr} 
= 0.34 \, {\rm Myr}$ for $e = 0.7$ and $q = 0.01$; and $a_{\rm cr} \simeq 
6 r_{\rm G}$ and $t_{\rm cr} = 2000 \, {\rm yr}$ if $e = 0$ and $q = 5
\times 10^{-5}$.

      When $a < a_{\rm cr}$, the inspiraling secondary black hole begins 
to push the inner disk inwards on a gravitational radiation time-scale 
(see also \citet{armitage02} for a circular system). When the 
semimajor axis $a$ is $\simeq n^{2/3} 2 r_{\rm G} \simeq 5.8 r_{\rm G}$,
the outer edge of the inner disk is at about $r_{\rm i} \simeq 
2 r_{\rm G}$ and the inner disk disappears. When $r_{\rm i} = 2 
r_{\rm G}$, the radial flow speed $v$ of the inner disk is the inspiral
speed of the secondary and $v = \dot{a}_{\rm gw} = - 1.6 \times 10^{-2} q 
\left(1 + q\right)^{1/2} f v_{\rm K2}$ where $v_{\rm K2}$ is the Newtonian
Kepplerian velocity at $2 r_{\rm G}$. For $q =0.01$ and $e = 0.7$, 
$v \simeq - 4.4 \times 10^{-3} v_{\rm K2} $. From Eq.~\ref{eq:avis}, 
$|\dot{a}_{\rm vis}| \sim 10^{-6} \delta_{-2}^2 \alpha_{-2} v_{\rm K}$ 
and $|v| \sim 10^{3} |\dot{a}_{\rm vis}|$. Thus, the dissipated energy 
will go into thermal energy instead of being radiated away and the inner 
disk becomes hotter and thicker \citep{begelman82}. If $\alpha$ 
does not change and $|v| \sim |\dot{a}_{\rm vis}|$, $\delta \simeq 
0.3$ and a thin disk assumption might be still valid. As $| v |
\ll v_{\rm K2}$ and the gap in the disk is determined by the dynamical
orbital resonance, it might be reasonable to assume that the size of the 
inner disk steadily reduces all the way from $r_{\rm i} \simeq n^{-2/3} 
a_{\rm cr}$ to $r_{\rm i} \simeq 2 r_{\rm G}$ due to the continuous push
of the secondary. The situation might be different from that in a binary
system of a circular orbit, in which a strong wind is suggested by  
\citet{armitage02}. In either case, it can be expected that 
the inner disk disappears around the time when the supermassive binary
black holes merge. 

      From Eq.~\ref{eq:agw}, the inspiraling secondary black hole 
evolves from $a_{\rm cr}$ to about $\sim r_{\rm G}$ on a time scale
$t_{\rm gw} \simeq t_{\rm cr} / 4$ if $e$ is constant. From 
Eqs.~\ref{eq:avis} and \ref{eq:acr}, the inner edge of the outer disk
moves inwards during $t_{\rm gw}$ from $r_{\rm o} \simeq 
n^{2/3} a_{\rm cr}$ to a radius 
\begin{equation}
 r_{\rm m} \simeq {a_{\rm cr} \over 4} \left(8 n - 3\right)^{2/3} .
\label{eq:ain}
\end{equation}
For the typical parameters of $\alpha = 0.01$, $\delta = 0.01$ and 
$M= 5 \times 10^8 {\rm M_\odot}$, we have $r_{\rm m} \simeq 310 r_{\rm G}$ for 
$q = 0.01$ and $e = 0.7$ ($n = 5$) and $r_{\rm m} \simeq 60 r_{\rm G}$
for $q = 5\times 10^{-5}$ and $e = 0$ ($n = 2$). Thus, when
the two supermassive black holes merge, a big hole ranging from 
$r_{\rm G}$ to $r_{\rm m}$ ($\gg r_{\rm G}$) forms in the inner  
disk around the post-merged black hole. 

      When the big hole forms, the accretion disk has no plasma to
fuel jets and jet formation stops. Jet formation revives only 
when the inner edge of the outer disk evolves from $r_{\rm m}$ to about
$2 r_{\rm G}$, which corresponds to $a <  n^{-2/3} 2 r_{\rm G} \simeq 
0.7 r_{\rm G}$. Therefore, the binary black holes must have 
merged before the jet formation restarts and the revival of 
jet formation ensures the coalescence of the supermassive binary
black holes. From Eq.~\ref{eq:avis}, the interruption time interval is
\begin{equation}
      t_{\rm m} \simeq {8 n -3 \over 12} t_{\rm cr} .
\label{eq:tint}
\end{equation}
For the typical super-massive binary system, $t_{\rm m} \simeq 
1.0\, {\rm Myr}$ for $ q = 10^{-2}$, $e=0.7$ and $n = 5$; and 
$t_{\rm m} \simeq 820\, {\rm yrs}$  for $ q = 10^{-3}$, $e=0$
and $n = 2$.

%----------------------------------------------------------------

\section[]{Interruption and restarting of jet formation in DDRGs}
\label{sec:appl}

\subsection[]{Interruption time scale}

%-----------------------------------------------------------------
\begin{figure}
\includegraphics[width=8.5cm]{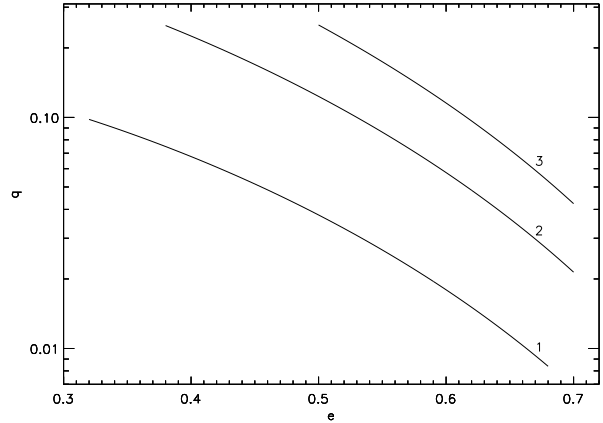}
 \caption{Relation between mass ratio $q$ and eccentricity $e$ for $\alpha 
= 0.01$ and $\delta =  H/ r = 0.01$. {\it Curve 1:} for central black hole
mass $M = 6 \times 10^8 {\rm M_\odot}$ and interruption time $t_{\rm m} =
1\times 10^6 \, {\rm yr}$ (the DDRG source B1834+620); {\it curve 2:} for
$M = 6 \times 10^7 {\rm M_\odot}$ and $t_{\rm m} = 2\times 10^5 \, {\rm 
yr}$; {\it curve 3:} for $M = 6 \times 10^6 {\rm M_\odot}$ and $t_{\rm m} 
= 3\times 10^4 \, {\rm yr}$.
}
\label{fig:qe}
\end{figure} 
%----------------------------------------------------------------

From the apparent magnitude $m_{\rm R} = 19.7$ of the host galaxy of the
DDRG source B1834+620 at red-shift $z = 0.5194$ \citep{schoenmakers00b}
and the correlation of central black hole mass and galaxy bulge 
luminosity for active and inactive galaxies \citep{mclure02}.
the central black hole mass is estimated to be $M \simeq 6 \times 10^8 \,
{\rm M_\odot}$. The observed interruption time of B1834+620 is $t_{\rm 
obs} = 1 \, {\rm Myr}$ \citep{schoenmakers00b} and from 
Eqs.~\ref{eq:tint} and \ref{eq:tcr} we have the interruption time
of jet formation 
\begin{eqnarray}
    t_{\rm m} & \simeq & 1.1  \delta_{-2}^{-16/5} \alpha_{-2}^{-8/5} 
         f_{(0.68)}^{3/5} \left({q \over 0.01}\right)^{3/5} \left(1 + 
	 q\right)^{-1/5} \times \nonumber \\
	     & & \left({M \over 6\times 10^8 M_\odot}\right) 
	 \, {\rm Myr}  ,
\label{eq:tb1834}
\end{eqnarray}
where $n = 5$ and $f_{(0.68)} = f/21.8$ for $e = 0.68$. For 
typical disk parameters $\alpha = 0.01$ and $\delta = 0.01$, 
Fig.~\ref{fig:qe} shows the relation of the eccentricity $e$ and the
mass ratio $q$ for B1834+620. For a circular orbit $e=0$,  $q > 1$
is required. Therefore, the orbit of the binary system in B1834+620
is not circular but elliptical. On the other hand, Fig.~\ref{fig:qe}
shows that if $q \ll 10^{-2}$, $e > 0.7$. \citet{artymowicz92} shows that 
when $e \ga 0.7$ a binary system suffers a slow eccentricity damping
for any $q$ and it is difficult to keep an extremely high eccentricity
$e \ga 0.8$ for a long time. In fact, the orbit of the secondary black 
would be circularized by binary-disk interaction for $q < 10^{-2}$  
\citep{artymowicz92}. Therefore, we conclude that the binary harbored in 
B1834+620 is elliptical with $10^{-2} \la q \ll 1$ and 
$0.3 \la e \la 0.7$. 

\begin{figure}
\includegraphics[width=8.5cm]{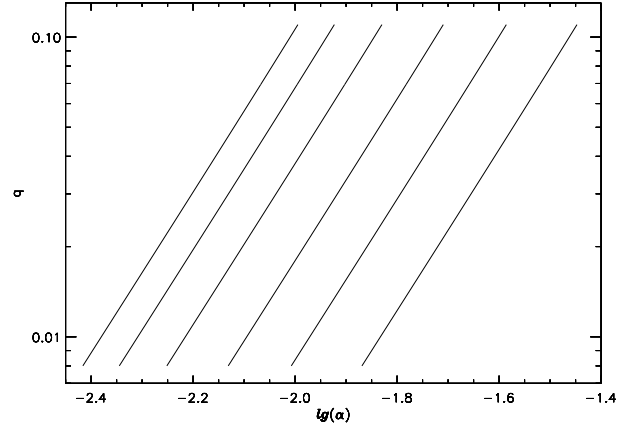}
 \caption{Mass ratio $q$ as a function of $\alpha$ for the DDRG source 
        B1834+620 with $M = 6 \times 10^8 M_\odot$, $t_{\rm m} = 1\, 
	{\rm Myr}$ and $\delta = 0.01$. The curves correspond, from left
	to right, to $e = 0.3$, $0.4$, 
	$0.5$, $0.6$, $0.68$, and $0.75$, respectively.
}
\label{fig:qalpha}
\end{figure}

      Eq.~\ref{eq:tb1834} indicates that the interruption time 
$t_{\rm m}$ is sensitive to disk parameters $\alpha$ and $\delta$. 
Fig.~\ref{fig:qalpha} gives $q$ as a function of $\alpha$ and $e$ for 
B1834+620. For $0.01 \la q \la 0.1$ and $0.3 \la e \la 0.75$, $\alpha$ is
in the range $3.5 \times 10^{-3} \la \alpha \la 4.0 \times 10^{-2}$.
If  $\delta = 0.01$ and $e = 0.68$, the mass of the secondary black 
hole in B1834+620 is $m \simeq 5 \times 10^6 {\rm M_\odot}$
for $\alpha = 0.01$;  $m \simeq 3 \times 10^7 {\rm M_\odot}$
for $\alpha = 0.02$; and $m \simeq 1 \times 10^8 {\rm M_\odot}$ for 
$\alpha = 0.03$, respectively. 

   As the central black hole masses in AGNs are in the ranges $10^{7.5} 
{\rm M_\odot} \la M \la 10^{9.5} {\rm M_\odot}$ (e.g. \citet{wu02}), 
the possible interruption time of jet formation 
is $50 \, {\rm Kyr} \la t_{\rm m} \la 5\, {\rm Myr}$, if the 
disk-binary system is typical with $\alpha = 0.02$, 
$\delta = 0.01$, $q = 0.05$, and $e = 0.68$. When the 
interruption time scale $t_{\rm m}$ is of order $ 10^6 \, {\rm yr}$, 
warm clouds of gas embedded in the hot intergalactic medium can fill
the old outer cocoon and the new jets may give rise to two new radio 
lobes in FR II radio galaxies \citep{kaiser00}, while if $t_{\rm 
m} \ll 10^6 \, {\rm yr}$ no new inner radio lobe can form by the 
restarting jets, like in 3C288 \citep{bridle89}, 3C219 \citep{clarke92},
and B0108+388 \citep{baum90,owsianik98}.

%-----------------------------------------------------------------
%-----------------------------------------------------------------

\subsection{FR II radio morphology and detection rate of DDRGs
}

One requirement for the massive secondary to open a gap in the 
accretion disk and to migrate inwards on viscous time at 
a large separation of the binary is  $M_{\rm d} \ga m$ \citep{syer95,
ivanov99}. The size of a thin standard 
accretion disk $r_{\rm d} \sim 10^4 r_{\rm G}$ may be determined
by star formation in the out-most regions of the disc or by the 
specific angular momentum of the gas which enters the disk. For a
simple $\alpha$-disk \citep{shakura73}, the steady-state 
disk surface density is given by
\begin{equation}
\Sigma \simeq 3.5\times 10^4 \alpha_{-2}^{-4/5} M_8^{1/5} \dot{m}_{-2}^{3/5}
              r_4^{-3/5} \, {\rm g\, cm^{-2}} ,
\label{eq:sigma}
\end{equation}
where $r_4 = r_{\rm d} / 10^4 r_{\rm G}$ and $\dot{m} = \dot{M} / 
\dot{M}_{\rm Edd} = 10^{-2} \dot{m}_{-2} $ with the Eddington 
accretion rate $\dot{M}_{\rm Edd}= 1.2 M_8 \, ({\rm M_\odot \, yr^{-1}})$. 
From Eq.~\ref{eq:sigma}, we have
\begin{equation}
M_{\rm d}/m = {10 \pi \Sigma r_{\rm d}^2 \over 7 m} \simeq 7
              \alpha_{-2}^{-4/5} M_8^{6/5} \dot{m}_{-2}^{3/5} r_4^{7/5} 
	      \left({q \over 0.05}\right)^{-1} .
\label{eq:ratiods}
\end{equation}
FR I and FR II radio galaxies can be separated clearly according to their
radio power \citep{fanaroff74} and/or to their optical luminosity 
of the host galaxy in the sense of increasing dividing radio luminosity 
with increasing optical luminosity of the host galaxy \citep{ledlow96}.
The division line in the radio power - host galaxy optical luminosity 
plane corresponds to a critical accretion rate \citep{ghisellini01}
\begin{equation}
\dot{m}_{\rm cr} = {\dot{M} \over  \dot{M}_{\rm Edd}} \sim 3 \times 10^{-2}
                  \left({\epsilon \over 0.2}\right)^{-1} .
\label{eq:crac}
\end{equation}
In FR I radio galaxies, the accretion rate $\dot{m} < \dot{m}_{\rm 
cr}$, while in FR II radio galaxies $\dot{m} > \dot{m}_{\rm cr}$. 
A low accretion rate $\dot{m} \la 10^{-2}$ in FR I radio galaxies and
a high accretion rate $\dot{m} \gg 10^{-2}$ in FR II radio 
galaxies are also suggested by \citet{cavaliere02,bottcher02}.
From Eqs.~\ref{eq:ratiods} and \ref{eq:crac}, 
$M_{\rm d}/m \ga 1 $ in FRII radio galaxies, while $M_{\rm d}/m \la 
1 $ in FR I radio galaxies. For an accretion disk with $\dot{m} \la 
10^{-2}$, the accretion does not appear in a thin or slim disk but 
possibly in ADAF \citep{narayan94,abramowicz95}, while
the accretion disk is geometrically thin \citep{shakura73} for
$10^{-2} \ll \dot{m} \la 1$ or slim \citep{abramowicz88} when $\dot{m} 
\ga 1$. Therefore, only the binary black holes in FR II or possibly  
boderline FRI/FRII radio galaxies can lead to the interruption
and restarting of jet formation and DDRGs should have FRII or 
borderline FRI/FRII radio morphology. As in the model the primary, the 
secondary and the accretion disk are roughly coplanar with one another,
the rotating post-merged supermassive black hole is thus roughly aligned
with the rotating primary and  the new-born jets in DDRGs should restart
symmetrically and roughly in the same directions of former jets. 

When the secondary migrates from $a_{\rm cr}$ to $r_{\rm G}$ and pushes the 
gas in the inner accretion disk inwards with a velocity $|\dot{a}_{\rm gw}| >
|\dot{a}_{\rm vis}|$ \citep{armitage02}, the accreting mass into
the primary black hole and down to the jets increases dramatically. The 
jets become extremely strong and the formed extremely large (giant) outer
lobes of DDRGs have relatively high luminosity 
as compared to the inner radio structure formed by the recurrent
jets. As the life time of a binary system in AGNs is very long
\citep{begelman80,quinlan97} and our model 
concerns the last stage of the supermassive binary black holes,
a DDRG source is a post-merged galaxy and should form a largest (giant)
possible structure of a few hundred Kpc and no possibility to find
clear evidence for galaxy merging in its host galaxy.

The time for jet material to travel from the central nuclei 
to the extended radio lobes in DDRGs is $\sim\, {\rm Myr}$ \citep{kaiser00}
and the possible time $t_{\rm DD}$ to detect a radio galaxy with 
a DDRG is the total time of the interruption ($\sim \, {\rm Myr}$) and the 
traveling time ($\sim \, {\rm Myr}$) of jet plasma from the central core
to radio lobe. If every FRII radio galaxy harbors a super-massive binary 
once in its life time $\sim 10^9\, {\rm yr}$ and we take  $t_{\rm DD} \sim 
10^7 \, {\rm yr}$, the possibility to detect a FRII radio source with a 
DDRG is $\sim$ one percents. This is consistent with the observations
of a low detection rate of DDRGs.

%%%%%%%%%%%%%%%%%%%%%%%%%%%%%%%%%%%%%%%%%%%%%%%%%%%%%%%%%%%%%%%
\section{Discussions and conclusions}
\label{sec:con}

We present a super-massive binary black hole scenario to explain the
interruption and restarting of jet formation in DDRGs. The orbit of 
the secondary with a mass ratio $10^{-2} \la q \ll 1 $ is elliptical 
with a typical eccentricity $e \sim 0.68$ and coplanar with the accretion 
disk. The secondary opens a gap in the accretion disk at large separation 
of the binary, while when the binary merges the gap extends and becomes a 
big hole in the vicinity of the post-merged black hole, leading to the
interruption of jet formation.  When the big hole is refilled with 
accreting plasma, jet formation restarts. Before the hole is filled, the
two black holes must have already merged. We show that the viscous time 
for accreting matter to refill the big hole is the observed interruption
time of jet formation in DDRGs. Prior to the merge of the binary, the 
accretion rate becomes extremely high and the produced jet is very
strong. Thus, the formed outer radio structures are very large (giant)
and with relatively high luminosity as compared to the inner structure
formed by the recurrent jets. As the merger of binary black holes does not 
change the direction of the spinning axis of the central supermassive
black hole, the inner radio structure should align with the outer 
lobes. We also show that accretion disk only in FRII or borderline 
FRI/FRII radio galaxies could strongly interact with the supermassive 
binary black holes and DDRGs should have FR II radio morphology.

The binary orbit in the model is elliptical. Gravitational wave emission
is very effective for eccentric binaries. A high eccentricity 
significantly shortens the evolutionary time-scale of the binary and 
enlarges the big hole as compared to that dug by a circular binary.
The interruption time of a circular orbit system is too short to explain
the observations of DDRGs. In the course of binary evolution, dynamical 
friction with stars in the cluster around the central black hole 
is unlikely to lead to a substantial increase of the eccentricity 
\citep{polnarev94}, but the eccentricity of a binary system 
changes with time due to the interaction of the disk to the secondary
\citep{artymowicz92}. For a secondary with 
$q \la 10^{-2}$, the orbit is circularized at the initial stage when 
the binary-disk system forms. For a massive secondary of $10^{-2} \la 
q \la 10^{-1}$, the eccentricity is excited if $e \la 0.7$ and gets 
slowly damped if $e > 0.7$ \citep{artymowicz92}. Therefore, the required
binary parameters of $10^{-2} \la q \ll 1$ and $ 0.3 \la 
e \la 0.7$ at $a \simeq a_{\rm cr}$ are quite reasonable and the 
typical values $q = 0.05$ and $e \simeq 0.68$ are favorite and 
close to the balance value $e \simeq 0.70$. When we estimated the 
gravitation radiation dynamic time $t_{\rm gw}$, we have assumed 
the eccentricity $e$ be constant. When the loss of the angular 
momentum due to the gravitational radiation becomes dominated, the 
eccentricity slowly decreases with time. However, Eqs.~\ref{eq:ain} 
and \ref{eq:tint} show that the size of the big hole in the 
disk and the interruption time of jet formation are mainly determined 
by the binary parameters at $a \simeq a_{\rm cr}$ and the assumption 
of constant $e$ does not significantly change the result.

In general, the binary orbital plane of the secondary initially inclines with 
respect to the disk plane, when the secondary passes through the star
cluster in the galactic disk and reaches a separation $a < r_{\rm d}$.
\citet{ivanov99} show that the inner part of the disk with radius smaller
than some alignment radius $r_{\rm al}$ ($ \ga a$) becomes twisted on
a short time scale $t_{\rm al1}$ and lies in the orbital plane if $\alpha
> \delta$. When the inner disk becomes coplanar with the orbital plane of 
the secondary, the rotating primary black hole is realigned with the twisted
inner accretion disk due to the Lense-Thirring drag \citep{scheuer96}
on a relatively short time scale $t_{\rm al2}$ when $10^3 r_{\rm G} 
\ll a < r_{\rm d}$ \citep{natarajan98}. At the same time, the 
orientation of the binary orbital plane slowly changes with time and has
a tendency to become vertical to the outer accretion disk on a much longer 
time scale $t_{\rm al3}$ \citep{ivanov99}. The time scale $t_{\rm al3}$
depends on $\alpha$ and accretion rate $\dot{M}$ for $q > 10^{-3}$. 
For an accretion disk with $\alpha \simeq 1$, the time scale $t_{\rm al3}$
with $t_{\rm al3} \gg t_{\rm al1} \sim t_{\rm al2}$ \citep{liu02c}
is much smaller than the life time $\sim 10^9\, {\rm yr}$ of an active 
galaxy, while for binary systems like those in DDRGs with $10^{-2} \la
q \ll 1$ and $\alpha \ll 1$, the situation is more complicated  
\citep{ivanov99,scheuer96}. But it is still 
possible for the orbital plane of the secondary to be coplanar with the
accretion disk within a reasonable time scale, as the vertical shear
of the twisted disk may be much stronger than its azimuthal counterpart 
\citep{papaloizou83,kumar85,natarajan98}. 

When the primary, the secondary and the accretion disk become coplanar 
with one another, the orientation of the spinning axis of the primary
dramatically change twice and so does the orientation of jet. The fist 
change happens on short time $\sim t_{\rm al2}$, while the second does
on a much longer time scale $ t_{\rm al3}$. When jets change their 
orientations,  X-shaped radio structure forms \citep{liu02c}.
As the rapid realignment happens only when the accretion disk is a thin 
$\alpha$-disk with $M_{\rm d} / m \ga 1$,  X-shaped radio structure, like 
double-double radio lobes in DDRGs, can be detected only in FR II or 
extremely luminous FR I radio galaxies. 
The detailed discussions on how our model works for the X-shaped radio 
galaxies \citep{dennett02} and what is the relation of the
X-shaped radio galaxies and the DDRGs is beyond the scope of the present 
paper and will be presented in a further work \citep{liu02c}.

When the semimajor axis of the orbit is smaller than the critical radius, 
the gap rapidly increases with the decrease of the separation. When the 
binary is close and almost ready to merge, the inner accretion disk 
becomes extremely hot and strong outflow might form. The sources may become
extremely bright in X-ray. The gravitational wave radiation of the binary
system is very strong and the system becomes a very good target for the 
monitoring of gravitational wave detectors. However, such strong X-ray and 
gravitational radiation sources may be  difficult to be discovered, as their
life time is less than a few thousand years.
When two supermassive black holes gets merged, the inner
region of accretion disk becomes empty and no X-ray and radio
radiation comes from the accretion disk and the jets. It is possible in 
a large sky survey to detect some sources with luminous radio lobes and 
bright jet-fragments but with a weak central nucleus in radio and X-ray
wavebands.

\section*{Acknowledgments}

We thank Prof. D.N.C. Lin for helpful comments and Dr. Xuelei Chen for 
interesting discussions. This work is supported by NSFC (No. 10203001)
and the Swedish Natural Science Research Council (NFR).

\bsp

\label{lastpage}

\end{document}